\documentclass{article}
\usepackage[utf8]{inputenc}
\usepackage{amsthm,amsmath,latexsym,amssymb,amsfonts,amscd}
\usepackage{authblk}
\usepackage[all]{xy}

\usepackage[all]{xy}
\newtheorem{theorem}{Theorem}[section]

\newtheorem{proposition}[theorem]{Proposition}

\theoremstyle{definition}

\theoremstyle{remark}

\newcommand{\mB}{\mathrm{B}}

\title{\bf What is the Seiberg--Witten map exactly?}
\author[1]{Vladislav \textsc{Kupriyanov}} 
\author[2]{Alexey \textsc{Sharapov}}
\affil[1] {\it Centro de Matem\'atica, Computa\c{c}\~{a}o e
Cogni\c{c}\~{a}o, Universidade Federal do ABC, Santo Andr\'e, SP, 
Brazil} 
\affil[2]{\it Physics Faculty, Tomsk State University, Lenin ave. 36, Tomsk 634050, Russia}

\date{}

\begin{document}

\maketitle

\begin{abstract}
We give a conceptual treatment of the Seiberg--Witten map as a quasi-isomorphism of $A_\infty$-algebras.
\end{abstract}
\section{Introduction}

Based on several nontrivial examples, the authors of Ref. \cite{madore2000gauge} suggested that {\it ``there should be an underlying geometric reason for the Seiberg--Witten map''}. This paper resulted from our attempt to find such a reason. The answer we have arrived at, however, is purely algebraic rather than geometric.  
Before going into details, let us recall the statement of the problem that the Seiberg--Witten  map solves.

In the simplest situation, one considers  a gauge vector field $\mathcal{A}$ on an $n$-dimensional Minkowski space with coordinates $x^i$. The standard Maxwell's equations for the gauge potential $\mathcal{A}$ are obtained by varying the action functional
\begin{equation}\label{SA}
    S[\mathcal{A}]=\int d^n x \mathcal{F}_{ij} \mathcal{F}^{ij}\,,
\end{equation}
where $\mathcal{F}_{ij}=\partial_i\mathcal{A}_j-\partial_j \mathcal{A}_i$ is the strength tensor of the electromagnetic field. 
The action is invariant under the usual gauge transformations 
\begin{equation}\label{gtr}
    \delta_{\lambda}\mathcal{A}_i=\partial_i \lambda\,,
\end{equation}
$\lambda$ being the gauge parameter. 

Suppose now that, in addition to the metric, the Minkowski space carries a constant bivector $\theta$. This allows one to endow the space of smooth functions of the $x$'s with the Weyl--Moyal $\ast$-product\footnote{As is customary for physical considerations we leave aside the question of convergence of the series.}:
\begin{equation}\label{WM}
    (f\ast g)(x)=\lim_{y\rightarrow x}e^{\frac12\theta^{ij}\frac{\partial}{\partial y^i}\frac{\partial}{\partial x^j}} f(y)g(x)\,.
\end{equation}
This product is known to be associative but not commutative unless $\theta = 0$. In particular,
$[x^i, x^j]_\ast=\theta^{ij}$. Hence, a nonzero $\theta$ brings in  noncommutativity in the space-time that the gauge vector field  lives on. 
In order to distinguish between commutative and noncommutative cases, we will `put hats' on fields living on the noncommutative Minkowski space. 
Following the pattern of non-Abelian gauge theories, it is now quite natural to replace the infinitesimal gauge transformation (\ref{gtr}) with
\begin{equation}\label{gtr1}
      \hat{\delta}_{\hat{\lambda}}  \hat {\mathcal{A}}_i=\partial_i \hat \lambda+[\hat{\mathcal{A}}_i, \hat {\lambda}]_\ast
      \end{equation}
and to define the gauge-covariant strength tensor by the familiar expression
\begin{equation}\label{F}
\hat{\mathcal{F}}_{ij}=\partial_i\hat{\mathcal{A}_j}-\partial_j\hat{\mathcal{A}_i}  +[\hat{\mathcal{A}}_i, \hat{\mathcal{A}}_j]_\ast \,.
\end{equation}
Then, $\delta_{\hat {\lambda}}\hat{\mathcal{F}}=[\hat{\mathcal{F}},\hat \lambda]_\ast$. Unlike (\ref{gtr}), the gauge transformations (\ref{gtr1}) do not commute to each other:
\begin{equation}
[\hat\delta_{\hat{\lambda}_1}, \hat\delta_{\hat{\lambda}_2}]=\hat\delta_{[\hat{\lambda}_1,\hat{\lambda}_2]_\ast}\,.
\end{equation}
Finally, the $\ast $-product algebra of functions enjoys the trace functional
\begin{equation}
    \mathrm{Tr}(f)=\int d^n x f(x)\,.
\end{equation}
The main property of the trace reads
\begin{equation}
    \mathrm{Tr}(f\ast g)= \mathrm{Tr}(g\ast f)=\int d^n x f(x)g(x)\,,
    \end{equation}
provided the functions $f$ and $g$ vanish at infinity fast enough.
With the trace, one can define the gauge invariant action $S[\hat{\mathcal{A}}]$  by the same integral (\ref{SA}) with $\mathcal{F}$ replaced by $\hat{\mathcal{F}}$. Because of the $\ast$-commutator in (\ref{F}), the corresponding equations of motion are not linear anymore. 

A remarkable fact established in \cite{Seiberg_1999} is that  being dynamically different, the commutative and noncommutative field theories have essentially the same gauge structure. More specifically, there exists a bijective mapping between the spaces of fields $\mathcal{A}$ and $\hat{\mathcal{A}}$ that takes gauge equivalent configurations of fields to gauge equivalent ones. This by no means implies an isomorphism of the corresponding gauge algebras, regarded as infinite-dimensional Lie algebras. (Such  an isomorphism is clearly impossible because one algebra is Abelian and the other is not.) It turns out that an appropriate geometric framework for understanding the above equivalence is that of Lie algebroids rather than Lie algebras.  Here we will not dwell on this issue in detail, referring the interested reader to the papers \cite{Barnich:2010xq} and \cite[Sec. 5]{Lyakhovich_2005}.  In a nutshell, one may allow the structure constants of the gauge algebra to depend on fields (though it is not our case). In this more general setting, it is legitimate to consider redefinitions of gauge fields that are accompanied by {\it field-dependent} redefinitions of gauge parameters. Such transformations form the natural automorphism group  of a gauge theory in the Batalin--Vilkovisky (BV) formalism. Regarding now a linear representation of a Lie algebra as a particular case of Lie algebroids, one can  consider the transformations
\begin{equation}\label{sw}
\hat{\mathcal{A}}=\hat{\mathcal{A}}(\mathcal{A})\,,\qquad \hat{\lambda}=\hat{\lambda}({\lambda, \mathcal{A}})\,.
\end{equation}
The condition that gauge equivalent field configurations go into gauge equivalent ones leads to the  equation $\delta_{{\lambda}}[\hat{\mathcal{A}}(\mathcal{A})]=\hat \delta_{\hat\lambda}\hat{\mathcal{A}}(\mathcal{A})$ or, more explicitly, 
\begin{equation}\label{sw1}
\hat{\mathcal{A}}_i (\mathcal{A}+\partial\lambda)-\hat{\mathcal{A}}_i(\mathcal{A})=\partial_i\hat{\lambda}(\lambda, \mathcal{A})  +[ \hat{\mathcal{A}}_i(\mathcal{A}),  \hat{\lambda}(\lambda,\mathcal{A})]_\ast\,,
\end{equation}
with infinitesimal $\lambda$ and $\hat\lambda$. Requiring further the gauge algebra to close off-shell, one also imposes the gauge consistency condition 
\begin{equation}\label{sw2}
    \delta_{\lambda_2}\hat \lambda_1-\delta_{\lambda_1}\hat \lambda_2=[\hat \lambda_1,\hat \lambda_2]_\ast-\hat{\lambda}([\lambda_1,\lambda_2]_\ast,\mathcal{A})\,,
    \end{equation}
    where the variation $\delta_{\lambda_2}\hat{\lambda}_1$ refers to the $\mathcal{A}$-dependence of $\hat{\lambda}(\lambda,\mathcal{A})$ and the gauge transformation (\ref{gtr}).
Geometrically, formulas (\ref{sw1}, \ref{sw2}) identify the transformation  (\ref{sw}) as a morphism of two Lie algebroids associated with the action of the Abelian and non-Abelian Lie algebras on the space of fields $\mathcal{A}$.  Any solution to Eqs. (\ref{sw1}, \ref{sw2}) with the initial condition 
\begin{equation}\label{bc}
    \hat{\mathcal{A}}=\mathcal{A} +O(\theta)\,,\qquad \hat{\lambda}=\lambda+O(\theta)
\end{equation}
is called a Seiberg--Witten (SW) map. 
The existence of SW maps  is not obvious in advance. Original arguments in its support came from string-theoretic considerations \cite{Seiberg_1999}. It is easy to check that the following expressions satisfy the above equations up to the first order in $\theta$:
\begin{equation}
\begin{array}{l}
   \displaystyle \hat{\mathcal{A}}_i=\mathcal{A}_i-\frac12\theta^{kj}\mathcal{A}_k(\partial_j\mathcal{A}_i+\mathcal{F}_{ji})+O(\theta^2)\,,\\[3mm]
 \displaystyle   \hat{\lambda}=\lambda+\frac{1}{2}\theta^{ij}\partial_i\lambda \mathcal{A}_j+O(\theta^2)\,.
    \end{array}
\end{equation}

In this paper, we give a refined algebraic explanation of the origin of the SW map, free from all inesential field-theoretic details.  More concretely, we show that any SW map defines and is defined by a quasi-isomorphism of differential graded algebras (dg-algebras for short). The dg-algebras underlying the (non)commutative gauge theories above appear to be very simple and are discussed in the next section. Certainly, it is not the first time the term `quasi-isomorphism' is mentioned  in connection with SW-maps. Thus, in \cite{Brace_2001, Cerchiai:2002ss, GBarnich_2001, Barnich_2004}, the problem of constructing SW maps was put in the framework of BV-BRST formalism.  In that approach, each gauge theory is described by a BV master action on the ghost-extended space of fields and antifields. Expanding the master action in powers of antifields generates a sequence of multilinear operators, which can be regarded as the structure maps of a cyclic $L_\infty$-algebra.  The defining relations of an $L_\infty$-algebra (a.k.a higher Jacobi identities) follow from the classical master equation. From this perspective, each SW map gives rise to a (quasi-)isomorphism of the corresponding  $L_\infty$-algebras. In  explicit terms, the relationship between SW maps and quasi-isomorphisms of $L_\infty$-algebras was explored in \cite{Blumenhagen_2018}. 

Our setup is somewhat different. Instead of $L_\infty$-algebras, we work in the category of $A_\infty$-algebras, which seems more natural, as the $\ast$-product algebra (\ref{WM}) responsible for noncommutativity is associative. Our construction of quasi-isomorphism in no way involves the equations of motion, which are the main (and unavoidable) ingredient of the BV-BRST formalism. This leads to further conceptual simplification. The $L_\infty$-algebras, their quasi-isomorphisms and, eventually, SW maps are obtained by the standard antisymmetrization procedure from the corresponding $A_\infty$-algebras and their morphisms, much as the commutator in an associative algebra yields a Lie bracket. The quasi-isomorphism construction is recursive and results in explicit expressions for the field transformation (\ref{sw}).  Since the recursion is on the degree of $\mathcal{A}$, the result appears to be nonperturbative in $\theta$, as opposed to the earlier approaches \cite{Seiberg_1999, madore2000gauge, Asakawa_1999, doi:10.1142/S201019451200685X}. The importance of nonperturbative in $\theta$ solutions (e.g. for studying instantons) was stressed in the original paper \cite{Seiberg_1999}. The algebraic treatment also provides the homotopy classification of ambiguities in SW maps, which we briefly address in Sec. 3.  
All in all, we hope that the new conceptual understanding gained in this paper will be useful for further generalizations and applications of the SW map. 
Some interesting possibilities are discussed in the concluding section.

Finally, a word about notation.  Throughout the paper, we work over the field of reals $\mathbb{R}$, 
though all our results are valid for any field of characteristic zero. All unadorned tensor products $\otimes$ and $\mathrm{Hom}$'s are defined over $\mathbb{R}$.  
When dealing with graded vector spaces and algebras, we tend to write down all sign factors explicitly. However, one can easily keep track of signs using the Koszul rule: whenever two graded objects $a$ and $B$ are swapped, the factor of $(-1)^{\deg a \cdot \deg B}$ appears.

\section{SW map as a quasi-isomorphism of dg-algebras} 
Let $A_\theta$ denote a unital dg-algebra over $\mathbb{R}$ generated by two sets of elements $x^i$ and $\xi^i$, $i=1,\ldots, n$, subject to the relations 
\begin{equation}\label{xx}
    x^ix^j-x^jx^i=\theta^{ij}\,,\qquad x^i\xi^j-\xi^j x^i=0\,,\qquad \xi^i\xi^j=0\,.
\end{equation}
Here $\theta=(\theta^{ij})$ is a real antisymmetric matrix. By definition,
\begin{equation}\label{deg}
    \deg x^i=0\,,\qquad \deg \xi^i=1
\end{equation}
and
\begin{equation}
    dx^i=\xi^i\,,\qquad d\xi^i=0\,.
\end{equation}
It follows from (\ref{xx}, \ref{deg}) that  $A_\theta$ is concentrated in degrees zero and one. In case $\theta=0$, the dg-algebra  $(A_0, d)$ is obviously commutative. For any $\theta$  the algebra $A_\theta$ satisfies the Poincar\'e--Birkhoff--Witt (PBW) property: as an $\mathbb{R}$-vector space it is isomorphic to the space of polynomials $a(x,\xi)$ that are at most linear in the $\xi$'s. So, the general element of $A_\theta $ is represented by a polynomial
\begin{equation}\label{aa}
    a=a_0(x)+a_i(x)\xi^i\,,
    \end{equation}
    where $a_0(x)$ and $a_i(x)$ are polynomials in the $x$'s. 
The product is then defined by the Weyl--Moyal type formula
\begin{equation}
    (f \ast_\theta g)(x,\xi)=\mu \sum_{n=0}^{\infty}\frac{1}{n!}\Big(\frac12\theta^{ij}\frac{\partial}{\partial x^i}\otimes \frac{\partial}{\partial x^j}\Big)^n f\otimes g \,,
\end{equation}
where
$$
\mu(f\otimes g)= p (fg)\,,
$$
and $p$ is the projection onto the subspace of polynomials at most linear in $\xi$'s. The polynomials (\ref{aa}) that are independent of the $\xi$'s form a subalgebra and those that are linear in  $\xi$'s constitute a nilpotent ideal.

The cohomology algebra $H(A_\theta)$ is obviously
commutative and does not depend on $\theta$, i.e., $H(A_\theta)\simeq H(A_{\theta'})$. In particular, $H^0(A_\theta)=\mathbb{R}$.

\begin{proposition}\label{p1}
For all antisymmetric matrices $\theta$ and $ \theta'$ the corresponding  dg-algebras $(A_\theta, d)$ and $(A_{\theta'}, d)$ are quasi-isomorphic. 
\end{proposition}

Recall that two dg-algebras $A$ and $B$ are said to be quasi-isomorphic if there exists a zig-zag sequence of dg-algebra morphisms 
$$
A\leftarrow C_1\rightarrow C_2\leftarrow C_3\rightarrow\cdots \leftarrow C_n\rightarrow B
$$
such that each morphism induces an isomorphism in cohomology. In particular, this means that $H(A)\simeq H(B)$. A dg-algebra $A$ is called formal if it is quasi-isomorphic to its algebra of cohomology $H(A)$.

To prove Proposition \ref{p1}, let us define the operator $d^\ast: A_\theta\rightarrow A_\theta$ of degree $-1$ that acts on cochains (\ref{aa}) as
\begin{equation}
    d^\ast a=x^ia_i(x)\,.
\end{equation}
Then the algebra $H(A_\theta)$ is isomorphic to the subalgebra $Z=\ker d\cap \ker d^\ast\subset A_\theta$. The subalgebra $Z$, being commutative, does not depend on $\theta$. Hence, we have the zig-zag map
\begin{equation} \label{AZA}
    A_\theta\stackrel{i}{\leftarrow}Z\stackrel{i}{\rightarrow} A_{\theta'}\,,
\end{equation}
where $i$ stands for the natural embedding. This proves Proposition \ref{p1}. We also see that the algebra $A_\theta$ is formal. 

It is known that whenever two dg-algebras  $A$ and $B$ are quasi-isomorphic they are quasi-isomorphic as $A_\infty$-algebras, meaning the existence of a single quasi-isomorphism $f: A\rightarrow B$. Let us recall some basic definitions. 

Given a graded vector space $A=\bigoplus A_n $, define the shifted graded space $A[1]=\bigoplus A_{n+1}$, that is, $A[1]_n=A_{n+1}$.  In order to avoid confusion we will denote the degree of a homogeneous element $a \in A[1]$ by bar:
\begin{equation}
    \bar a=\deg a-1\,.
\end{equation}
(On the right, $a$ is treated as an element of $A$.)

The bar coalgebra of $A$ is given now by the direct sum 
\begin{equation}
    B(A)=\bigoplus_{l=1}^\infty B^l(A)\,,\qquad B^l(A)=A[1]^{\otimes l}\,,
\end{equation}
endowed with the coproduct 
\begin{equation}
    \Delta (a_1\otimes \cdots \otimes a_n)=\sum_{l=1}^{n-1} (a_1\otimes \cdots\otimes a_l)\bigotimes (a_{l+1}\otimes \cdots \otimes a_n)\in B(A)\bigotimes B(A)\,.
\end{equation}
Recall that a coderivation of a coalgebra $C$ is a linear map $D: C\rightarrow C$ such that 
\begin{equation}
    (D\otimes 1+1\otimes D)\Delta a=\Delta (Da) \qquad \forall a\in C\,.
\end{equation}
The coderivations form a graded Lie algebra $\mathrm{Coder}(C)$ w.r.t. the graded commutator.
As is well known, the space of coderivations of the coalgebra $B(A)$ is isomorphic to the space 
$$
\mathrm{Hom}(B(A), A[1])=\bigoplus_{k=-\infty}^\infty\mathrm{Hom}^k(B(A), A[1])\,, 
$$
so that to every  $D\in \mathrm{Hom}(B^l(A), A[1])$ there corresponds the coderivation of the form
\begin{equation}
    \hat D (a_1\otimes\cdots\otimes a_n) =\sum_{i=0}^{n-l}\pm  a_1\otimes \cdots \otimes a_i\otimes D(a_{i+1},\ldots,a_{i+l})\otimes \cdots \otimes a_n\,,
\end{equation}
$\pm$ being the Koszul sign. A coderivation $\delta$ of degree one is called a codifferential if $\delta^2=0$. 
By definition, an $A_\infty$-algebra structure on a graded vector space $A$ is a codifferential on $B(A)$. 
In other words, an $A_\infty$-algebra is defined by an element $m\in \mathrm{Hom}^1(B(A), A[1])$ such that $\hat m\hat m=0$. Expanding $m$ in homogeneous components, $m=m_1+m_2+m_3+\cdots$, we get a sequence of multilinear maps $m_l\in \mathrm{Hom}^1(B^l(A), A[1])$ obeying the quadratic relations
\begin{equation}
    \hat m_1\hat m_1=0\,, \qquad \hat m_1\hat m_2+\hat m_2\hat m_1=0\,,\qquad \ldots
\end{equation}
In particular, for any dg-algebra $A$ one can set $m_1(a)= da$, $m_2(a,b)=(-1)^{\bar a}ab$, and $m_l=0$ for all $l>2$.  Hence, each dg-algebra is an $A_\infty$-algebra for which only the two maps $m_1$ and $m_2$ may be different from zero. 

If $(A, m)$ and $(A', m')$ are $A_\infty$-algebras, then a morphism between them is a morphism of differential graded coalgebras $F: B(A)\rightarrow B(A')$. The morphism $F$ is completely defined by a map $f\in \mathrm{Hom}^0(B(A), A'[1])$, often called twisting cochain,  obeying the relation
\begin{equation}
    m'_1(f)+m'_2(f,f)+m'_3(f,f,f)+\cdots = f\circ \hat{m}\,.
\end{equation}
For dg-algebras $(A, d)$ and $(A', d')$ this amounts to the sequence of equations 
\begin{equation}\label{QC}
  \delta f_l=  \Psi_l[f_1,\ldots, f_{l-1}] \,,\qquad l=1,2,\ldots,
\end{equation}
where 
$$
    (\delta f_l)(a_1,\ldots,a_l)=d' f_l(a_1,\ldots, a_l)-\sum_{i=1}^l (-1)^{\bar a_1+\cdots+ \bar a_{i-1}}f_l(a_1,\ldots, da_i,\ldots, a_l)\,,
$$
\begin{equation}\label{psi}
\begin{array}{c}
    \Psi_l[f_1,\ldots, f_{l-1}](a_1,\ldots, a_l)= \displaystyle \sum_{i=1}^{l-1} (-1)^{\bar a_1+\cdots+\bar a_i}f_{l-1}(a_1,\ldots, a_i a_{i+1},\ldots a_l)\\[5mm]
-\displaystyle \sum_{i=1}^{l-1} (-1)^{\bar a_1+\cdots +\bar a_i}f_i (a_1,\ldots,a_i)f_{l-i} (a_{i+1},\ldots, a_{l})\,,
\end{array}
\end{equation}
and $f_l$ stands for the restriction of $f$ onto the subspace $B^l(A)$.
For $l=1$, $\Psi_1\equiv 0$ and  $f_1: A[1]\rightarrow A'[1]$
defines a cochain transformation, i.e., $d'f_1=f_1 d$. 
{We say that a cochain transformation $f_1$ is {\it unital} if it preserves the unit, that is, $f_1(1)=1$. }

A morphism $F$ is called quasi-isomorphism if $f_1$ induces an isomorphism $H(f_1): H(A)\rightarrow H(A')$ in cohomology. 

Notice that the operator $\delta$ defines a differential that makes each space $\mathrm{Hom}(B^l(A), A'[1])$ into a cochain complex. By induction on $l$, one can see that $\delta \Psi_l=0$; and hence, all $\Psi_l$'s are $\delta$-cocycles. The quasi-isomorphism condition (\ref{QC}) requires  all these cocycles to be trivial. We conclude that whenever the differential $\delta$ is acyclic in degree $1$ any morphism of complexes $f_1: A\rightarrow A'$ extends to a morphism 
of $A_\infty$-algebras. Of course, the acyclicity of $\delta$ is only a sufficient (not necessary) condition for the existence of an $A_\infty$-morphism $f$.

We say that a cochain $f\in \mathrm{Hom}(B^l(A), A'[1])$   is {\it normal}  if it satisfies the normalization condition: $f(a_1,\ldots,a_l)=0$ whenever any of the arguments is equal to $1\in A$. Obviously, the normal cochains constitute a subcomplex of the complex $\mathrm{Hom}(B^l(A), A'[1])$. 

\begin{proposition}\label{p2}
If $f_1$ is unital and $f_2, f_3,\ldots, f_{l-1}$ are normal, then the cochain $\Psi_l[f_1,\ldots,f_{l-1}]$ defined by Eq. (\ref{psi}) is normal as well. 
\end{proposition}
The proof is by direct verification.

Let us specify the above formulas for the case of algebras $A=A_\theta$ and $A'=A_{\theta'}$. As both the algebras are nested in degrees zero and one, $A_\theta=(A_\theta)_0\oplus (A_\theta)_1$, each map $f_l$ splits into $l+1$ components 
\begin{equation}
\begin{array}{l}
    \displaystyle   f_l':\; (A_\theta)_{1}^{\otimes l}\rightarrow (A_{\theta'})_1\,,  \\[5mm]
   \displaystyle    f''_{l,k}: (A_\theta)_1^{\otimes k}\otimes (A_\theta)_0\otimes (A_\theta)_1^{\otimes (l-k-1)}\rightarrow (A_{\theta'})_0,\qquad k=0,1,\ldots,l-1.
\end{array}
\end{equation}
In terms of these component maps,  equations (\ref{QC}) decompose into two groups: 
\begin{equation}\label{r1}
\begin{array}{c}
 f'_l(a_1,\ldots, a_{k},db,\ldots, a_{l-1})=
d f''_{l,k}(a_1,\ldots, a_k, b, \ldots, a_{l-1})\\[3mm]
-f'_{l-1}(a_1,\ldots, a_{k}\ast_\theta b, \ldots, a_{l-1})+f'_{l-1}(a_1,\ldots, b\ast_\theta a_{k+1}, \ldots, a_{l-1})\\[3mm]
    +\displaystyle \sum_{n=1}^{k} f'_n(a_1,\ldots, a_n) \ast_{\theta'}f''_{l-n,k-n}(a_{n+1}, \ldots, a_k, b,\ldots, a_{l-1})\\ [5mm]
    -\displaystyle \sum_{n=k+1}^{l-1} f''_{n,k}(a_1,\ldots, a_k, b,\ldots,  a_{n-1})\ast_{\theta'} f'_{l-n}(a_{n}, \ldots, a_{l-1} )
    \end{array}
\end{equation}
for all $0\leq k\leq l-1$ and
\begin{equation}\label{r2}
\begin{array}{c}
    f''_{l,s}(a_1,\ldots, db_1, \ldots, a_{s-1}, b_2,\ldots, a_{l-2}) - f''_{l,k}(a_1,\ldots, a_k, b_1, \ldots, db_2,\ldots, a_{l-2}) \\[3mm]
    =-f''_{l-1,s-1}(a_1,\ldots,a_{k}\ast_\theta b_1,\ldots, a_{s-1}, b_2,\ldots, a_{l-2} )\\[3mm]
    +f''_{l-1,s-1}(a_1,\ldots,b_1 \ast_\theta a_{k+1},\ldots,a_{s-1}, b_2,\ldots, a_{l-2} )\\[3mm]
    +f''_{l-1,k}(a_1,\ldots,a_k, b_1,\ldots, a_{s-1}\ast_\theta b_2,\ldots, a_{l-2})\\[3mm]
    -f''_{l-1,k}(a_1,\ldots,a_k,b_1,\ldots, b_2\ast_\theta a_{s},\ldots, a_{l-2} )\\[3mm]
    +\delta_{k, s-1}f''_{l-1,k}(a_1,\ldots,a_k, b_1 \ast_\theta b_2, a_s, \ldots, a_{l-2} )  \\[0mm] 
    \displaystyle-\sum_{n=k}^{s-1}f''_{n+1,k}(a_1,\ldots, a_k, b_1,\ldots, a_n)\ast_{\theta'} f''_{l-n-1,s-n-1}(a_{n+1},\ldots, b_2,\ldots, a_{l-2}) 
     \end{array}
\end{equation}
for all $0\leq k\leq s-1\leq l-2$. As is seen the second group of equations  does not involve the $f'$'s, giving thus a closed subsystem for determining unknown maps $f''_{k,l}$.  We will present the general solution to Eqs. (\ref{r1}, \ref{r2}) in the next section. For now, let us explain the relevance of $A_\infty$-morphisms above to SW maps. 
Given elements $a\in (A_\theta)_1$ and $b\in (A_\theta)_0$, define  the pair of chains 
\begin{equation}\label{fp}
    \varphi=\sum_{l=1}^\infty a^{\otimes l}\,,\qquad \psi=\sum_{s,u=0}^\infty a^{\otimes s}\otimes b\otimes a^{\otimes u}
\end{equation}
of the bar coalgebra $B(A_\theta)$. Clearly, $\bar \varphi=0$ and $\bar \psi=1$. Let $\hat D_a$ denote the coderivation of $B(A_\theta)$ associated with the cochain $D_a\in \mathrm{Hom}(B^1(A_\theta), A_\theta[1])$ defined by 
\begin{equation}\label{Da}
D_a(c)=dc+a\ast_\theta c - c\ast_\theta a \qquad \forall c\in A_\theta[1] \,.
\end{equation}
Then evaluating  equation  (\ref{r1}) on $\psi$ and summing up over $l$, one can see that the cochains 
\begin{equation}
f' = \sum_{l=1}^\infty f'_l\,,\qquad f''=\sum_{l=1}^\infty\sum_{k=0}^{l-1} f''_{l,k}    
\end{equation}
of $\mathrm{Hom}(B(A_\theta), A_{\theta'}[1])$ satisfy the equation 
\begin{equation}\label{ffD}
d f''(\psi)+f'(\varphi)\ast_{\theta'}f''(\psi)-f''(\psi)\ast_{\theta'}f'(\varphi) =f'(\hat D_a\psi)\,.
\end{equation}
Now it remains to make the following identifications:
\begin{equation}
    a\leftrightarrow \mathcal{A}\,,\qquad b\leftrightarrow \lambda\,,\qquad f'(\varphi)\leftrightarrow \hat{\mathcal{A}}(\mathcal{A})\,,\qquad f''(\psi) \leftrightarrow     \hat{\lambda}(\lambda,\mathcal{A})\,.
\end{equation}
With such identifications, Eq.(\ref{ffD}) reproduces the equivalence condition (\ref{sw1}) if one set $\theta=0$ and then change $\theta'\rightarrow \theta$. Thus,  we  get the conclusion that the SW map is nothing but a quasi-isomorphism $f$ of the dg-algebras $A_0$ and $A_{\theta}$ in the category of $A_\infty$-algebras. Similarly, evaluating equations (\ref{r2}) on the chain 
\begin{equation}\label{chi}
    \chi =\sum_{s, u, v=0}^\infty a^{\otimes s}\otimes \big(b_1\otimes a^{\otimes u}\otimes b_2- b_2\otimes a^{\otimes u}\otimes b_1\big)\otimes  a^{\otimes v}\,,
\end{equation}
one can deduce the identity 
\begin{equation}\label{gff}
    -f''(\hat D_a \chi)=f''(\psi_1)\ast_{\theta'}f''(\psi_2) -f''(\psi_2)\ast_{\theta'}f''(\psi_1) -f''(\psi_{12})\,,
\end{equation}
where 
\begin{equation}\label{psis}
\begin{array}{l}
   \displaystyle \psi_i=\sum_{s,u=0}^\infty a^{\otimes s}\otimes b_i\otimes a^{\otimes u}\,,\quad i=1,2\,,\\[5mm]    \displaystyle \psi_{12}=\sum_{s,u=0}^\infty a^{\otimes s}\otimes (b_1\ast_{\theta} b_2-b_2\ast_{\theta} b_1)\otimes a^{\otimes u}\,.
    \end{array}
    \end{equation}
Upon identification  $b_i \leftrightarrow \lambda_i$, Rel. (\ref{gff}) reproduces the gauge consistency condition (\ref{sw2}).

Note that all the chains (\ref{fp}, \ref{chi}, \ref{psis}) are totally symmetric in permutations of tensor factors. This means that we evaluate the multilinear maps $f_l$ on the symmetric diagonal. Upon symmetrization, the maps $f_l$ define an $L_\infty$-morphism between $A_\theta$ and $A_{\theta'}$ treated as dg-Lie algebras w.r.t. the commutators.

\section{Explicit solutions and ambiguity}
As a preparatory step, we introduce a complex that controls the solution space. 
First, one can regard $B^l(A_\theta)$ as the $l$th tensor power of the complex $A_\theta[1]$ with the differential induced by $d$. 
Let $C^l$ denote the corresponding cochain complex with coefficients in $A_{\theta'}[1]$. By definition, 
\begin{equation}
C^l=\mathrm{Hom}(B^l(A_\theta), A_{\theta'}[1])=\bigoplus_{k=-1}^l \mathrm{Hom}^k(B^l(A_\theta), A_{\theta'}[1])\,.
\end{equation}
The second equality is due to the fact that the graded space $B^l(A_\theta)$ is concentrated in degrees $0, -1,\ldots, -l$, while $A_{\theta'}[1]$ is nested in degrees $0$ and $-1$. We denote the differential in $C^l$ by $d_l$, that is, $ d_l (f)=f \circ d$ for all $f\in C^l$.  
The rationale behind the 
introduction of this sequence of complexes  is that the left- and right-hand sides of Eqs. (\ref{r1}, \ref{r2}) define, respectively, coboundaries and cocycles of the complex $C^l$ in degree one. As above by $\bar C^l$ we denote the subcomplex of normal cochains.

It is known that any operator in the space
of polynomials can be written as a differential operator (possibly of infinite order) with polynomial coefficients, see e.g. \cite[Lem. 5.1]{pinczon1997equivalence}. This allows us to identify the cochains of ${C}^l$ with polydifferential operators. Each polydifferential operator, in its turn, is completely determined by its symbol. The differential $d_l$  makes  then the space of symbols into a cochain complex, which we are going to describe in more detail below. 

Considering the dg-algebra $A_\theta$  just as a complex, denote by $\mB$ the complex dual to $A_\theta$.
By definition, the cochains of $\mB=\mB_{-1}\oplus \mB_0$ are given by the sums 
\begin{equation}\label{phi}
    \phi=\phi^0(p)+\phi^i(p)\zeta_i\,,
\end{equation}
where $\phi^0$ and $\phi^i$ are formal power series in the commuting indeterminates $p_1,\ldots, p_n$ and we prescribe the following degrees to the formal variables:
\begin{equation}
    \deg p_i=0\,,\qquad \deg \zeta_i=-1\,.
\end{equation}
The dual differential $\partial: \mB\rightarrow \mB$ increases the degree by one  and acts on  cochains (\ref{phi}) by the rule
\begin{equation}
    \partial \phi= p_i \phi^i(p)\,.
\end{equation}
The complex $(\mB, \partial)$ enjoys the natural descending  filtration $F^k B\supset F^{k+1}B$ by powers of the $p$'s. Define the subspace $Z=Z_{-1}\oplus Z_0$ of  $\mB$ as
\begin{equation}\label{ZZ}
Z_{-1}=\big\{\phi=\phi^i(p)\zeta_i\;\big |\; p_i\phi^i(p)=0\big\}\,, \qquad Z_0=\big\{\phi=c \in \mathbb{R}\big\} \,.
\end{equation}
The nonzero elements of $Z$ are obviously nontrivial $\partial$-cocycles and $H(\mB)\simeq Z$.

Considering now the space $A_{\theta}$ as a complex with zero differential, we define the tensor product of complexes 
\begin{equation}\label{Bk}
   \mB^l= A_{\theta}\otimes \mB^{\otimes l}\,.
\end{equation}
Let $\partial_l$ denote the coboundary operator in $\mB^l$. It is convenient to define the degree of a homogeneous cochain 
\begin{equation}\label{Phi}
\Phi=a\otimes \phi_1\otimes \cdots \otimes \phi_l\in \mB^l
\end{equation} 
to be equal to  $$\deg \Phi=\deg \phi_1+\cdots+\deg \phi_l+\deg a +l-1\,.$$
For this grading, the complex $\mB^l$ is  concentrated in degrees $-1,0,1,\ldots, l$. 

Since we are working  with complexes of $\mathbb{R}$-vector spaces, the  K\"unneth formula yields immediately 
\begin{equation}
    H(\mB^l)=A_{\theta}\otimes H(\mB)^{\otimes l}\,.
\end{equation}

For each cochain  (\ref{phi}) one can assign the map $\sigma(\phi): A_\theta\rightarrow (A_\theta)_0$ that takes (\ref{aa}) to 
$$
\sigma( \phi)(a)=\phi^0(\partial_i)a_0(x)+\phi^j(\partial_i) a_j(x)\,,
$$
where $\partial_i=\partial/\partial x^i$.  Similarly, to each cochain (\ref{Phi})
one can assign the multilinear operator of $C^k$ by setting 
\begin{equation}\label{sF}
    \Sigma(\Phi)(a_1,\ldots, a_l)=a \cdot \sigma(\phi_1)(a_1) \cdot \sigma(\phi_2)(a_2)\cdots \sigma(\phi_l)(a_l)\,.
\end{equation}
Here the dot stands for the usual product of polynomials and the result is treated as an element of $A_{\theta'}$. 
This assignment extends  to the whole of $\mB^l$ by linearity.

The filtration in $\mB$ induces that in the complex $\mB^l$ and we use the bold-font  letter $ \mathbf{B}^l$ to denote the completion of $\mB^l$ w.r.t. this filtration. The map $\Phi\mapsto \Sigma(\Phi)$ extends then to ${\mathbf{ B}}^l$ and we arrive at the following
\begin{proposition}
The map $\Sigma: \mathbf{B}^l\rightarrow C^l$ is an isomorphism of complexes, i.e., $d_l \Sigma=\Sigma \partial_l$.
\end{proposition}

It is the elements of  $\mathbf{B}^l$ which are identified with the symbols of  operators. 
The isomorphism $\Sigma$ allows one to transfer the notion of normality from $C^l$ to $\mathbf{B}^l$.  Clearly, 
the cochain (\ref{sF}) is normal iff  $\phi_1^0(0)=\cdots=\phi^0_l(0)=0$.  The normal cochains of $\mathbf {B}^l$ form a subcomplex, which we denote by $ \bar{\bf B}^l$. The formula
\begin{equation}
    f(p)-f(0)=p_i\int_0^1d t \left(\frac{\partial f}{\partial p_i}\right)(t p)\,,
\end{equation}
valid for any formal power series $f(p)$, shows  that each normal zero-chain $\phi=\phi^0(p)\in \mB^1_0$ is a coboundary of $\mB^1$. It follows from the K\"unneth formula  that 
\begin{equation}\label{HB}
    H^n(\Bar{ \mB}^l)=\left\{
    \begin{array}{l}
    (A_\theta)_0\otimes Z_{-1}^{\otimes l}\,,\quad n=-1\,;\\[3mm]
      (A_\theta)_1\otimes Z_{-1}^{\otimes l}\,,\quad n=0\,; \\[3mm]
        0\,,\quad n>0\,.\\[3mm]      \end{array}\right.
\end{equation}
Here $Z_{-1}$ stands for the  subspace of cocycles (\ref{ZZ}). 
Since the differential $\partial_l$ is acyclic in positive degrees when restricted to the subcomplex of normal cochains $\bar {\mathbf{B}}^l$,  so is the differential $d_l$ in the isomorphic complex $\bar C^l$. In particular, $H^{1}(\bar C^l)=0$. This means the absence of  obstructions to the solubility  of Eqs. (\ref{r1}, \ref{r2}). 

To write down an explicit expression for $f$ we need to choose an appropriate  homotopy operator that relates the identity map with a projection onto the subspace of nontrivial cocycles (\ref{HB}). First, we introduce the projector $\pi: B\rightarrow B$ onto the subspace $Z\subset B$ defined by (\ref{ZZ}).  For every (\ref{phi}), we set  
$$
\pi \phi=\phi^0(0)+\phi^i(p)\zeta_i-\zeta_j\frac{\partial}{\partial p_j}\int_0^1 p_i\phi^i(t p) d t\,.
$$
One can see that $\pi^2=\pi$ and $\mathrm{Im} \,\pi=Z$. Furthermore, $\pi$ is homotopic to the identity map. The corresponding homotopy operator $h: B\rightarrow B$ is given  by 
\begin{equation}\label{h}
    h\phi=\zeta_i\int_0^1\left(\frac{\partial \phi ^0}{\partial p_i}\right)(t p)d t\,.
\end{equation}
It is straightforward to check that 
\begin{equation}
    \partial h+h\partial =1-\pi
\end{equation}
and 
\begin{equation}
    h^2=0\,,\qquad \pi h=0\,,\qquad h\pi=0\,.
\end{equation}

Next, we extend the maps $\pi$ and $h$ from the complex $B$ to the tensor product  $B^l$ by the standard formulas of homological algebra. For any cochain (\ref{Phi}), we set
\begin{equation}
    \pi_l (\Phi)=a_0\otimes \pi\phi_1\otimes\cdots\otimes \pi\phi_l\,,
\end{equation}
\begin{equation}
    h_l (\Phi)=\sum_{i=1}^l a_0\otimes \phi_1\otimes \cdots \otimes \phi_{i-1}\otimes h\phi_i\otimes \pi \phi_{i+1}\otimes\cdots\otimes \pi \phi_l\,.
\end{equation}
Then
\begin{equation}\label{rel1}
    \partial_l h_l+h_l\partial_l =1-\pi_l\,,
\end{equation}
and
\begin{equation}\label{rel2}
    h_l^2=0\,,\qquad \pi_l h_l=0\,,\qquad h_l\pi_l=0\,,\qquad \pi_l^2=\pi_l\,.
\end{equation}
The isomorphism $\Sigma$ allows one to push forward the operators $\pi_l$ and $h_l$ from $\mathbf{B}^l$ to $C^l$. This results in 
the operators $h^\Sigma_l=\Sigma h_l\Sigma^{-1}$ and $\pi^\Sigma_l=\Sigma \pi_l\Sigma^{-1}$, which satisfy the same Rels. (\ref{rel1}, \ref{rel2}).

Let us set $f''_l=\sum_{k=0}^{l-1}f''_{l,k}$ and rewrite Eqs. (\ref{r1}, \ref{r2}) in the form
\begin{equation}\label{sys}
    d_l f'_l=\Theta'_l [f'_1,\ldots, f'_{l-1};f''_1,\ldots, f''_{l}]\,,\qquad d_lf''_l=\Theta''_k[f''_1,\ldots, f_{l-1}'']\,.
\end{equation}
Here the $\Theta$'s  are determined by the r.h.s. of  Eqs. (\ref{r1}, \ref{r2}). 

Since $\pi_l^\Sigma \Theta_l'=\pi_l^\Sigma \Theta_l''=0$, applying the homotopy operator $h^\Sigma_l$ brings  system (\ref{sys}) into the sequence of recurrence relations that determine the morphism $f$:
\begin{equation}\label{rr}
\begin{array}{l}
        f''_l=h^\Sigma_l\Theta''_l[f''_1,\ldots,f_{l-1}'']\,,\qquad l=2,3,\ldots ;\\[3mm]
        f'_m= h^\Sigma_l\Theta'_l[f'_1,\ldots,f'_{m-1};f''_1,\ldots,f''_m]\,,\qquad m=1,2,\ldots
        \end{array}
\end{equation}
As the initial value for these recurrence relations, one takes a unital cochain $f''_1$. 
It follows from  Proposition \ref{p2} that the cochains  $\Theta''_l $ and $\Theta'_l$ are normal whenever so are the cochains $f''$'s and $f'$'s they are built 
from. Since the operator $h^\Sigma_l$ respects normality, the resulting cochains $f''_l$ and $f'_l$ are also normal.  Thus, by induction, we see that relations (\ref{rr}) are well-defined provided $f''_1(1)=1$. Since $H^0(A_\theta)\simeq \mathbb{R}$, the last condition  implies that $f_1$ induces an isomorphism $H(f_1): H^0(A_\theta)\rightarrow H^0(A_{\theta'})$ of the zero cohomology groups.  To be a quasi-isomorphism,  $f$ must also generate a cohomology isomorphism in degree one. The simplest way to satisfy the quasi-isomorphism condition is to take $f''_1$ to be  the identity map on the space of polynomials $a_0(x)$, see (\ref{aa}). Then $f'_1=h_1^\Sigma \Theta_1'=h_1^\Sigma df''_1$ is again the identity map on the polynomials $a_i(x)\xi^i$. On the other hand, the identity map $f_1=(f_1',f''_1)=\mathrm{id}$, being an isomorphism of complexes, trivially satisfies Eq. (\ref{r1}) for $l=1$.   Therefore, $f_1=\mathrm{id}$ can always be chosen as an initial value for the recurrence relations (\ref{rr}). 
Exactly this choice corresponds to the classical Seiberg--Witten map (\ref{sw}, \ref{bc}). With $f_1=\mathrm{id}$, the total $A_\infty$-morphism $f$ is obviously invertible and we conclude that the algebras $(A_\theta, d)$ and $(A_{\theta'}, d)$, regarded as $A_\infty$-algebras, are isomorphic to each other.

In principle, one could try to construct an $A_\infty$-morphism whose component maps $f_l$ with $l>1$ are not normal  or $f_1$ is not  unital. However, under such weak assumptions on the $f_l$'s, the solubility of Eqs. (\ref{r1}, \ref{r2}) gets out of control because all obstruction spaces $H^1(C^l)\neq 0$. {Therefore, the existence of such `abnormal' $A_\infty$-morphisms is not clear at the moment}. 

It remains to discuss the arbitrariness in the solutions of Eqs. (\ref{r1}, \ref{r2}) under the assumption of normality.  Since, at each order $l$, we have a system of inhomogeneous linear equations $\delta f_l=\Psi_l$ for the known $\Psi_l$ and unknown $f_l$,  the solution space is isomorphic  to the space of normal $\delta$-cocycles in degree zero. In other words, it is possible to shift a solution  $f_l=(f'_l,f''_l)$ by any $\delta$-cocycle $g_l =(g_l', g_l'')\in \bar C^l$. According to (\ref{HB}), one can always choose $g_l'=z_l+ dc_l$ and $g_l''=d_lc_l$, where $\Sigma^{-1}(z_l)$ belongs to the completion of the space $(A_\theta)_1\otimes Z_{-1}^{\otimes l}$, while $c_l$ is an arbitrary homomorphism from $(A_\theta)_1^{\otimes l}$ to $(A_{\theta'})_0$. In particular, two cochain transformations $f_1, \tilde{f}_1: A_\theta\rightarrow A_{\theta'}$, considered up to homotopy, differ from each other  by an element $[z_1]\in H^0(\bar C^1)$.  
In the context of SW maps,  these ambiguities have been discussed in many physical papers, see e.g. \cite{Asakawa_1999, Barnich_2004, Muthukumar_2015, doi:10.1142/S201019451200685X, Aschieri_2019, Kupriyanov:2022ohu} and the references therein.  Since there is some  confusion in the literature about which ambiguities should be considered trivial and which are not, we will dwell on this issue in more detail.

  Isolating all the terms that are linear in $f$'s on the left  and summing up over $l$, we can write the sequence of equations (\ref{QC})  as a single Maurer--Cartan (MC) equation, namely,   
  \begin{equation}\label{MCE}
      Df =-\frac12[f, f]\,,\qquad f=\sum_{l=1}^\infty f_l\,.
  \end{equation}
Here the differential $D$ on the left involves $\delta$ and the product in $A$, while the commutator in $A'$ gives the Lie bracket on the right. The differential and the bracket make the space 
$\mathrm{Hom}(B(A), A'[1])[1]$ into a dg-Lie algebra $\mathcal{L}=\bigoplus\mathcal{L}_n$. The algebra $\mathcal{L}$ has the descending filtration $F^k\mathcal{L}\supset F^{k+1}\mathcal{L}$ by the number of arguments of homogeneous cochains. In particular,
\begin{equation}
    \sum_{l=n}^\infty f_l\in F^n\mathcal{L}\,.
\end{equation}
Due to the shift in degree,  the morphism $f$ gets degree one and defines an MC element of  $\mathcal{L}_1$.  Since $\mathcal{L}=F^1\mathcal{L}$, one can  formally exponentiate the Lie algebra $\mathcal{L}_0$ to produce the corresponding Lie group $G=\exp (\mathcal{L}_0)$. This group acts on the space of MC elements by the  following `gauge transformations' \cite{PMIHES_1988__67__43_0}:
\begin{equation}\label{heq}
f\quad \mapsto \quad f^g=g f g^{-1} -g^{-1}D g\qquad\quad \forall g\in G\,.
\end{equation}
Two MC elements (= morphisms) $f$ and $\tilde f$ are said to be gauge (= homotopy) equivalent if there exists an element $g\in G$ such that $\tilde f=f^g$.  In view of (\ref{heq}) it makes sense to consider the solutions to the MC equation (\ref{MCE}) up to gauge equivalence.  
Suppose now that two morphisms $f$ and $\tilde f$ coincide up to the order $l-1$ in filtration degree, i.e., $\tilde{f}-f\in F^{l}\mathcal{L}_1$.  Then it follows from (\ref{QC}) that $\delta(\tilde f_{l}-f_{l})=0$. If the $\delta$-cocycle $g_l=\tilde f_{l}-f_{l}$ happens to be trivial, that is $g_l=\delta c_l$, then  one can perform the gauge transformation $f\mapsto f^g$ with $g=\exp (c_l)$.  It is readily seen that $\tilde f-f^g\in F^{l+1}\mathcal{L}$; and  hence, the morphisms $\tilde f$ and $f^g$ coincide up to the order $l$. We thus conclude that the first nontrivial class of $\delta$-cohomology $\big[ \tilde f_{l}-f_{l} \big]$ represents an obstruction to the homotopy equivalence of two $A_\infty$-morphisms and the corresponding SW maps. A more systematic exposition of the homotopy theory of $A_\infty$ and other homotopy algebras can be found in \cite{dolgushev2015homotopy}.

\section{Discussion}
We presented an algebraic  treatment of the Seiberg--Witten map as a quasi-isomorphism of $A_\infty$-algebras. While in the original physical setting, the very existence of the SW map requires more or less nontrivial reasoning or calculations, it becomes a simple consequence of the dg-algebra quasi-isomorphism (\ref{AZA}) in the algebraic interpretation. A simple homotopy operator (\ref{h}) allows one to write down the corresponding SW map up to any given order in the gauge field $\mathcal{A}$. The simplicity of the  quasi-isomorphism  construction owes its existence  to two facts:
\begin{itemize}
    \item[(A)] The dg-algebras under consideration are isomorphic as cochain complexes and differ only in multiplication;
    \item[(B)] Every nontrivial cocycle has a central representative (homologically equivalent to an element from the centre $Z(A)$ of the algebra $A$).    
\end{itemize}
 Together, these  properties  provide the quasi-isomorphism $A\leftarrow Z\rightarrow A'$ with $Z=\ker d\cap Z(A)$.

There are many dg-algebras satisfying conditions A and B. For instance, one can replace the last relation in (\ref{xx})  with the standard anticommutativity condition $\xi^i\xi^j-\xi^j\xi^i=0$ for the odd generators.    The result may be viewed as a noncommutative version of the de Rham complex with $dx^i=\xi^i$. Let us denote it by $\Omega_\theta$. The ordinary (commutative) de Rham complex corresponds to $\theta=0$. Any two dg-algebras $\Omega_\theta$ and $\Omega_{\theta'}$  obviously satisfy the conditions above; hence, they are quasi-isomorphic. Furthermore, the dg-algebra $\Omega_\theta$ admits the natural descending filtration $F^{k}\Omega_\theta\supset F^{k+1}\Omega_\theta$ by the powers  of the $\xi$'s. This allows one to define the sequence of dg-algebras $A^{_{(k)}}_\theta=\Omega_\theta/F^k\Omega_\theta$. The dg-algebra with relations (\ref{xx}) corresponds to $k=2$. It is clear that any pair of dg-algebras from the family $A^{_{(k)}}_\theta$ satisfies conditions A and B whenever $k>1$. (For $k=1$, $H(A_\theta^{_{(1)}})\simeq A_\theta^{_{(1)}}$ and condition B is violated unless $\theta=0$.) 
The physical interpretation of the algebras $A_\theta^{_{(k)}}$ with $k>2$ is not clear  at the moment, except for the case $k=3$. As was shown in \cite{Blumenhagen_2018}, one can identify the additional homogeneous subspace $(A_\theta^{_{(3)}})_2$ in degree $2$ with the space of equations of motion (e.g. $\mathcal{F}=0$ as in Chern--Simons theory). Then the extra components of a quasi-isomorphism $f$ with arguments in $(A_\theta^{_{(3)}})_2$ allow one to cover the case of `open gauge algebra' when the gauge consistency condition (\ref{sw2}) involves an extra term vanishing on-shell.  

All the above constructions extend immediately to the tensor product of $A^{_{(k)}}_\theta$ with the matrix algebra $\mathrm{Mat}_N$.  Although the cohomology algebra of $A^{_{(k)}}_\theta \otimes \mathrm{Mat}_N$ is no longer Abelian, the subalgebra of cocycles $Z$ is still independent  of $\theta$, which is sufficient for the quasi-isomorphism. Upon replacing $\theta\rightarrow i\theta$, one can also reduce the matrix factor $\mathrm{Mat}_N$ to the Lie subalgebra of anti-Hermitian matrices $u(N)$. More general matrix factors were considered in \cite{JSSW, JMSSW, Chaich}. 

Another class of relevant dg-algebras is provided by the Wess--Zumino (WZ) complexes \cite{Wess:1990vh}. The corresponding dg-algebras are generated by the same formal variables $x^i$ and $\xi^i=dx^i$ but the relations on them are essentially different.  For the general discussion of WZ complexes, we refer the reader to \cite{manin1992notes, demidov1993some}. By way of illustration, we present the following (not most general) commutation relations: 
\begin{equation}\label{q}
\begin{array}{ll}
    x^ix^j-q^{ji}x^jx^i=0\,,&\quad \xi^i x^j-q^{ji}x^j\xi^i=0 \,,\\[3mm]
    \xi^i\xi^j+q^{ji}\xi^j\xi^i=0\,,&\quad (\xi^i)^2=0\,.
    \end{array}
\end{equation}
Here $q^{ij}$ is a set of real numbers subject to the condition $q^{ji}q^{ij}=q^{ii}=1$.
The dg-algebra generated by these $x$'s and $\xi$'s, let us denote it by $\Omega_{q}$, enjoys the PBW property and is isomorphic, as a vector space, to the usual de Rham complex. Moreover, one can see that the Poincar\'e Lemma holds true for $\Omega_q$, meaning $H(\Omega_q)\simeq \mathbb{R}\subset \Omega_q$. This gives the quasi-isomorphism $\Omega_q\leftarrow \mathbb{R}\rightarrow\Omega_{q'}$ of dg-algebras. Again, the natural grading on $\Omega_{q}$ by powers of the $\xi$'s gives rise to the descending filtration  $F^k\Omega_{q}\supset F^{k+1}\Omega_{q}$, so that one can define the sequence of dg-algebras $A^{_{(k)}}_{q}=\Omega_{q}/F^k\Omega_q$. The algebra $A^{_{(1)}}_q$, generated by the $x$'s, is often called the algebra of  `quantum polynomials'. Accordingly, the vector space spanned by the $x$'s is referred to as the `quantum space'.  
The next algebra,  $A^{_{(2)}}_{q}$, is the quantum counterpart of the algebra (\ref{xx}) underlying the Weyl--Moyal noncommutativity.  As with $A^{_{(2)}}_\theta$, the cocycles of $A^{_{(2)}}_{q}$ form a commutative algebra; hence, all the dg-algebras of the family $A^{_{(2)}}_{q}$ are quasi-isomorphic to each other. It should be mentioned that noncommutative field theory on quantum spaces was considered in \cite{madore2000gauge}, albeit with completely different differential and commutation relations among the $x$'s and $\xi$'s. Some examples of noncommutative gauge theories related to WS complexes have been elaborated in the recent paper \cite{MT}.

\section*{Acknowledgements}

We are grateful to Vasily Dolgushev and Maxim Grigoriev for useful discussions and references. This work was partially supported by the São Paulo Research Foundation (FAPESP), grants 2021/09313-8 and 2022/13596-8, and by the CNPq grant 304130/2021-4. A. Sh. acknowledges the financial support from the Foundation for the Advancement of Theoretical Physics and Mathematics ``BASIS''.

\end{document}